\documentclass[
prd,
twocolumn,
showpacs,superscriptaddress,groupedaddress
]{revtex4}
\usepackage[dvipdfmx]{graphicx}
\usepackage{microtype}
\usepackage{amsmath}
\usepackage{amssymb}
\usepackage{subfigure}
\usepackage{hyperref}
\usepackage{float,color} 
\usepackage{url}
\usepackage{xcolor}
\usepackage{color}
\usepackage{mathrsfs}
\usepackage{amsfonts}
\usepackage{latexsym}
\usepackage{ragged2e}
\usepackage{epsfig}
\usepackage{textcomp}
\usepackage{phaistos}
\usepackage{citing}

\makeatletter
\renewcommand\@makefnmark{\hbox{\@textsuperscript{\normalfont\color{purple}\@thefnmark}}}
\renewcommand\@makefntext[1]{%
  \parindent 1em\noindent
            \hb@xt@1.8em{%
                \hss\@textsuperscript{\normalfont\@thefnmark}}#1}
\makeatother

\usepackage{caption}
\DeclareCaptionJustification{justified}{\leftskip=0pt \rightskip=0pt \parfillskip=0pt plus 1fil}
\def\beq{\begin{equation}}
\def\eeq{\end{equation}}

\def\mathbb{\Bbb}

\newcommand{\alf}{Alfv\'en\ }
\newcommand{\alfv}{Alfv\'enic\ }
\newcommand{\al}{\alpha}
\newcommand{\be}{\beta}

\newcommand{\vp}{\varphi}
\newcommand{\del}{\delta }

\newcommand{\pa}{\partial}
\newcommand{\df}{\dfrac}

\newcommand{\mk}[1]{\left( #1 \right)}  
\newcommand{\kk}[1]{\left[ #1 \right]}

\definecolor{vividviolet}{rgb}{0.62, 0.0, 1.0}
\definecolor{amaranth}{rgb}{0.9, 0.17, 0.31}
\definecolor{palatinateblue}{rgb}{0.15, 0.23, 0.89}
\definecolor{brightpink}{rgb}{1.0, 0.0, 0.5}
\definecolor{cornflowerblue}{rgb}{0.39, 0.58, 0.93}
\definecolor{deepcarminepink}{rgb}{0.94, 0.19, 0.22}
\definecolor{radicalred}{rgb}{1.0, 0.21, 0.37}
\colorlet{RED}{red}

\hypersetup{ linktoc=all,
    colorlinks, linkcolor={palatinateblue},
    citecolor={brightpink}, urlcolor={amaranth}
}

\graphicspath{{Images/}}

\graphicspath{{Images/}}

\makeatletter
\def\@fnsymbol#1{\ensuremath{\ifcase#1
\or $\textleaf$ \or $\PHplaneTree$ \or $\PHrosette$ \or $\PHvine$
\else\@ctrerr\fi}}%

\makeatother

\def\sideremark#1{\ifvmode\leavevmode\fi\vadjust{\vbox to0pt{\vss
 \hbox to 0pt{\hskip\hsize\hskip1em
 \vbox{\hsize1.5cm\tiny\raggedright\pretolerance10000
 \noindent #1\hfill}\hss}\vbox to8pt{\vfil}\vss}}}%
\begin{document} 
\title{Unified understanding of \alfv superradiance and the Blandford-Znajek process in a force-free black hole magnetosphere}
%
\author{Sousuke \surname{Noda}}
\email{snoda@cc.miyakonojo-nct.ac.jp}
\affiliation{National Institute of Technology, Miyakonojo College, Miyakonojo, Miyazaki 885-8567, Japan}

 \author{Masaaki \surname{Takahashi}}
 \email{mtakahas@auecc.aichi-edu.ac.jp}
\affiliation{Department of Physics and Astronomy, Aichi University of
  Education, Kariya, Aichi 448-8542, Japan}

\begin{abstract}
We investigate the extraction of energy from a Kerr black hole via \alf waves (i.e., \alfv superradiance) in a force-free magnetosphere, in which the plasma inertia effects are ignored. 
We analyze the Poynting flux generated by \alf waves that propagate toward the event horizon across the inner light surface, the causal boundary for the waves. We find the relationship between the energy flux associated with \alf waves and that of the Blandford-Znajek (BZ) process. That is, both mechanisms can be described within a unified formulation of the Poynting flux, where the BZ process can be 
regarded as the long wavelength limit of the \alfv superradiance, and depending on the wave's frequency, \alf waves enhance or suppress the Poynting flux in the BZ process. This unified framework for the BZ process and the \alfv superradiance would offer a valuable perspective for understanding the energy sources of high-energy astrophysical phenomena, such as relativistic jets.

 \end{abstract}
      \pacs{04.70.Bw, 04.20.-q, 04.30.Nk, 52.35.Bj} 
      \keywords{Blandford-Znajek solution, superradiance for \alf waves, force-free electromagnetism}

\maketitle

\section{Introduction}

 In many highly energetic phenomena such as relativistic jets and gamma-ray bursts, the energy sources of the activities would be related to a black hole surrounded by plasma and 
 electromagnetic fields, forming a magnetosphere. In these phenomena, it is considered that 
 the energy can be extracted from the rotating black hole; e.g., by the Penrose process, superradiance of waves, and the Blandford-Znajek (BZ) process \cite{Blandford1977}. In particular, recent general relativistic magnetohydrodynamic (GRMHD) simulations for black hole 
 magnetospheres and comparisons with observations have widely suggested that the BZ process 
 extracts a huge amount of energy from the black hole \cite{McKinney2004,Komissarov2005,Tchekhovskoy2011,Narayan2012,Kino2022}.
 
 The energy extraction by the BZ process occurs due to the magnetic torque 
 induced by the spacetime dragging, which transfers the black hole's rotational energy into 
 the magnetosphere as the Poynting flux. 
 The original formulation of the BZ process is considered a stationary rotating force-free magnetosphere, allowing the effect of the plasma inertia to be neglected, and is derived as the conditions necessary for the outgoing 
 energy flux from the black hole horizon. In actual astrophysical environments, 
 however, black hole magnetospheres are not strictly 
 stationary, and force-free approximation may not be entirely valid. 
 Consequently, corrections such as those caused by the non-stationarity and plasma inertia effects to the BZ's energy extraction formula should be considered to achieve 
 a more comprehensive understanding of the energy 
 extraction from the black hole via magnetosphere. 

A correction to the energy flux was proposed by the authors \cite{Noda:2019mzd,Noda:2021ivs}. 
In these studies, we investigated the propagation of the 
\alf waves along a magnetic field line, by applying a perturbation to a force-free magnetosphere. It was found that the condition for the amplification 
of the \alf wave (i.e., \alfv superradiance) matches the condition for the BZ process to operate: $0<\Omega_F < \Omega_\text{H}$, where $\Omega_F$ and $\Omega_\text{H}$ denote the angular velocities of magnetic field lines and black hole horizon, respectively. However, this condition alone is not sufficient for concluding that \alf waves are able to extract the rotational energy. A fundamental problem with the condition identified in \cite{Noda:2019mzd,Noda:2021ivs} 
is that it does not depend on the \alf wave frequency, $\omega$.
This point is important since, in the energy extraction process through other kinds of waves (superradiance \cite{Zeldovich1971,Zeldovich1972,Starobinsky:1973aij,Starobinskil:1974nkd,Brito:2015oca}), 
the frequency plays a decisive role in determining whether energy is extracted or not.

 In this paper, we resolve the aforementioned concerns by deriving the $\omega$-dependence for the energy extraction condition via \alf waves propagating in a black hole magnetosphere within the force-free approximation. In the magnetosphere, there exists a special radius known as the inner light surface, where 
 the four velocity of an observer corotating with a magnetic field line becomes null. Inside this surface, the velocity exceeds the speed of light. Hence, \alf waves must propagate only in one direction along the magnetic field line. 
 By imposing a purely ingoing boundary condition for \alf waves at the inner light surface and evaluating the associated energy and angular momentum fluxes, we derive the $\omega$-dependence of the condition for energy extraction, namely the superradiant condition for \alf waves.
Notably, the functional form of the Poynting flux generated by \alf wave propagation has a common factor with that of the BZ process. 
This indicates that \alfv superradiance provides the wave-induced correction to the BZ power.

A key aspect of the wave-induced correction is that \alf waves pass through the inner light surface and fall into the black hole. This scenario generally holds for other configurations of magnetosphere with off-equatorial magnetic field lines. However, for the sake of simplicity and to 
clarify the essence of the wave effect on the BZ power, this study focuses on magnetic field lines on the equatorial plane of the Kerr spacetime.

The rest of the paper is organized as follows. In Sec.~II, we first present the background force-free magnetosphere near the equatorial plane of a Kerr black hole. 
We then give the \alf wave equation by applying a perturbation to the magnetosphere. In Sec.~III, we derive the condition for energy extraction via \alf waves. We also point out that the determinism associated with the threshold of 
this condition resembles that of negative-energy inflow in magnetohydrodynamics \cite{Takahashi1990}. In Sec.~IV, we show the Poynting flux generated by \alf waves and that from the BZ process are factorized with a common factor, and we discuss how \alf waves modify the BZ power through the wave effect. 
Sec.~V is devoted to concluding remarks.
Throughout the present paper, we measure the black hole mass $M$ and spin parameter $a$ in the unit of $c^2/G$ and $c$, respectively, so that they have dimensions of length.

\section{Force-free magnetosphere of Kerr black holes and \alf wave propagation}
\subsection{Force-free magnetosphere in the Kerr spacetime}
To study the \alf wave propagation, we consider the vicinity of the equatorial plane of a black hole magnetosphere. We use the Boyer-Lindquist coordinates 
$(t,r,\theta,\varphi)$ of the Kerr spacetime, of which the 
line element $ds^2=g_{\mu\nu}dx^\mu dx^\nu$ is given by
\begin{align}
  \notag ds^2&=
	-\left(1-\df{2M r}{\Sigma}\right)dt^2-\df{4aMr \sin^2{\theta}}{\Sigma}dt d\vp \\
	&\ \ \ \ +\dfrac{\Sigma}{\Delta}dr^2+\Sigma d\theta^2+\df{A \sin^2{\theta}}{\Sigma}d\vp^2,
         \label{eq:kerr}
 \end{align}
 where $\Delta=r^2-2Mr+a^2$\ , \ $\Sigma=r^2+a^2\cos^2{\theta}$\ ,\
   $A=(r^2+a^2)^2-\Delta a^2 \sin^2{\theta}$, and the constants $M$ and $a$ are the mass 
   and angular momentum per unit mass of the Kerr black hole, respectively. 
   The horizon radius $r_{\text{H}}$ is given as the larger root of $\Delta=0$. 
  The frame dragging effect is represented by the angular velocity of the zero angular 
  momentum observer (ZAMO), given by 
  $\Omega(r)=-g_{t\vp}/g_{\vp\vp}$.
At the horizon, the angular velocity becomes $\Omega _\text{H}:= a/(2Mr_\text{H})$.

When the electromagnetic fields are so strong that the plasma inertia effects can be ignored approximately, 
we obtain the condition on four current $j^\mu$ and field strength $F_{\mu\nu}$ as $F_{\mu\nu}j^{\nu}\approx 0$. 
This approximation leads to the following set of 
equations for $F_{\mu\nu}$:\ $ F_{\mu\nu}\nabla_{\alpha}F^{\nu\alpha}=0,
\ \nabla_{[\mu}F_{\nu\lambda]}=0$. The field strength satisfying these equations is expressed with two scalar potentials, $\phi_1$ and $\phi_2$, called Euler potentials as $F=d\phi_1 \wedge d \phi_2$ \cite{Uchida1}. Substituting this expression of the field strength into $F_{\mu\nu}\nabla_\al F^{\nu \al}=0$, we obtain 
\beq
   \partial_{\nu}\left[\sqrt{-g}\partial_{\mu}\phi_i\left(\partial^{\mu}\phi_1\partial^{\nu}\phi_2-\partial^{\nu}\phi_1\partial^{\mu}\phi_2\right)\right]=0,\ \ \ \ (i=1, 2).
        \label{eq:basicEuler}
\eeq
For axisymmetric stationary force-free fields, 
the physical meaning of Euler potentials is as follows \cite{Uchida2}:
$\phi_1(r,\theta)=\text{const}$ represents a magnetic surface on which magnetic field lines lie. 
On the other hand, $\phi_2=\text{const}$ describes the spatial 
configuration and time evolution of a magnetic field line on the magnetic surface.
A solution of \eqref{eq:basicEuler} in the vicinity of the equatorial plane of the Kerr spacetime is derived in 
\cite{Noda:2021ivs} as
 \begin{equation}
       \phi_{1}=q \cos{\theta},\ \phi_{2}=\varphi-\Omega_F t  + J_B \int \df{\Sigma}{\Delta} dr,\ \text{with}\ \left|\frac{\pi}{2} - \theta\right| \ll 1,
 \label{eq:EulerKerr}
 \end{equation} 
where $\Omega_F$ is the angular 
velocity of magnetic field lines on a magnetic. Note that $\Omega_F$ is regarded as a free parameter here \footnote{The value of $\Omega_F$, which is determined by 
outer boundary conditions, depends on the astrophysical environment around the black hole.  
In more realistic situations, the magnetic field would connect to plasma sources located somewhere between the inner and outer light surfaces. Therefore, as a phenomenological approach, we treat $\Omega_F$ as a free parameter so that the outer boundary condition can be specified by hand, depending on the astrophysical context.}.
Here, $\phi_1$ represents the monopole magnetic field with magnetic charge $q$. Using the regularity condition of $F_{\mu\nu}F^{\mu\nu}$ at the horizon, we have $J_B=(r_\text{H}^2+a^2)(\Omega_\text{H}-\Omega_F)/r_\text{H}^2$.

The four velocity of an observer corotating with a magnetic field line is given by $\xi_{(t)}^\mu+\Omega_F \xi^{\mu}_{(\vp)}$ with the timelike Killing vector $\xi_{(t)}^\mu$ and spacelike Killing vector $\xi_{(\vp)}^\mu$ of the Kerr spacetime. We denote the norm of this four velocity by $\Gamma=g_{tt}+2g_{t\vp}\Omega_F + g_{\vp\vp}\Omega_F^2$. 
According to \cite{Noda:2021ivs}, this can be factorized as follows $\Gamma=-\Omega_F^2(r_0-r)(r-r_\text{in})(r-r_\text{out})/r$,
where $r_0<0<r_\text{in} <r_\text{out}$. The two positive roots correspond to the radii of the inner and outer light surfaces. 
The inner and outer light surfaces are the causal boundaries for \alf waves.

\subsection{\alf wave propagation}
\begin{figure}[H]
 \centering
 \includegraphics[width=0.9\linewidth]{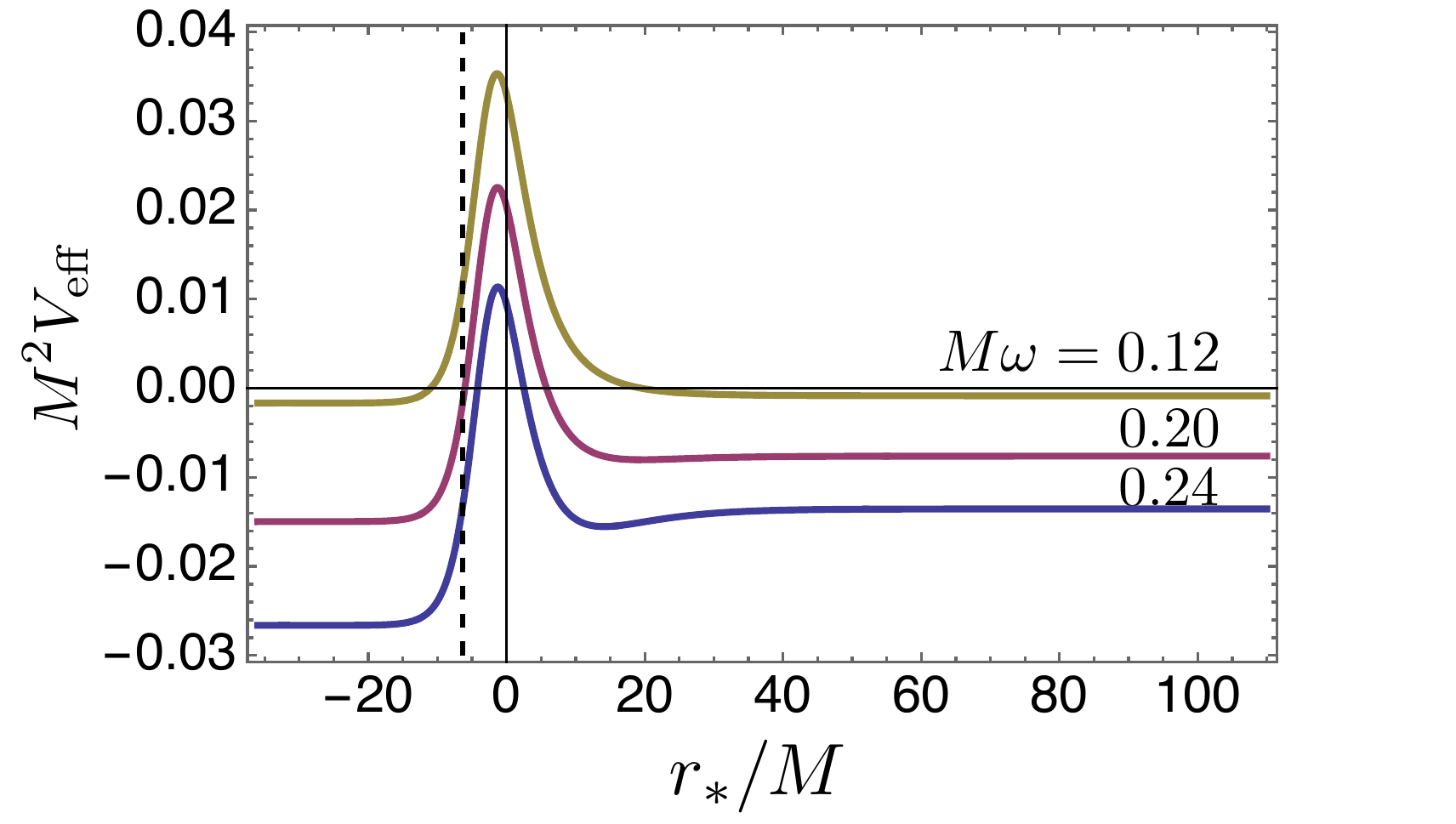}
\caption{\footnotesize{The effective potential for \alf waves $V_\text{eff}$. The parameters are fixed as $a/M=0.5, M\Omega_F=0.08$. As the frequency $\omega$ increases, the effective potential shifts downward. For this parameter set, the ergosurface is located at $r_*/M=-6.35$, which is shown as the vertical dashed line. Therefore, the potential peak lies outside the ergoregion.}}
\label{fig:potential}
\end{figure}

\begin{figure}[H]
 \centering
 \includegraphics[width=1\linewidth]{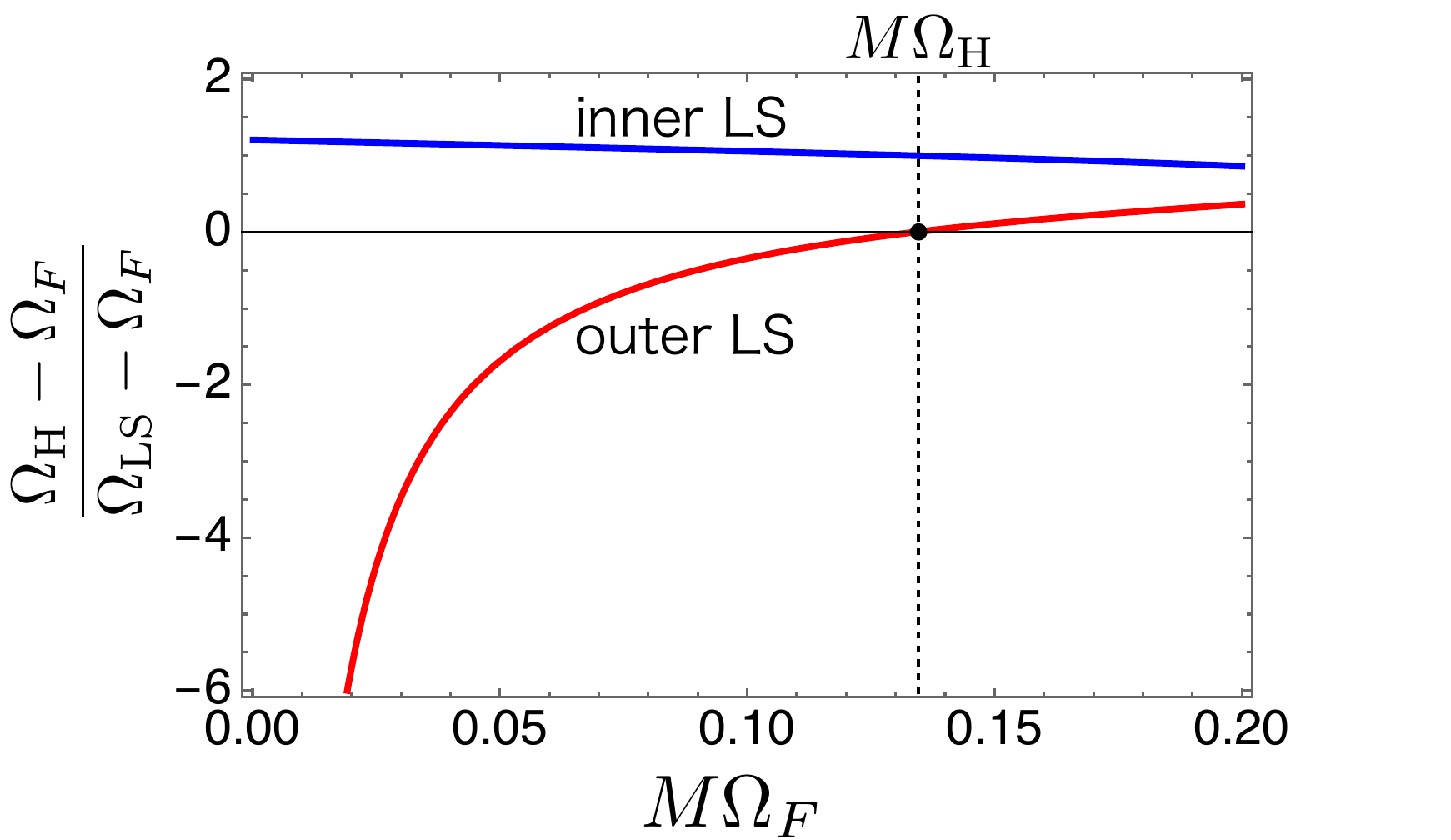}
\caption{\footnotesize{The $\Omega_F$-dependence on $(\Omega_\text{H}-\Omega_F)/(\Omega_{\text{LS}}-\Omega_F)$ for $a/M=0.5$. The function $\Omega_{\text{LS}}$ is $\Omega_{\text{in}}$ (blue curve) or $\Omega_{\text{out}}$ (red curve). In the parameter 
range $0<M\Omega_F < M\Omega_\text{H}$, the BZ process works.}}
\label{fig:phases}
\end{figure}

We apply a perturbation perpendicular to the magnetic surface lying on the equatorial plane, and it leads to the deviation of $\phi_1$ as $\phi_1 \rightarrow \phi_1 + \delta \phi$. 
Note that this specific perturbation does not induce any 
deviation in $\phi_2$, which corresponds to the fast magnetosonic wave due to the symmetry of the background magnetosphere.
The wave function $\delta \phi$ represents \alf waves, of which propagation is governed by the following equation \cite{Noda:2021ivs}:
\beq
\pa_\mu \left(\sqrt{-g}|\pa \phi_2|^2 P^{\mu\nu}\pa_\nu \delta \phi \right)=0,
\label{equatorial_waveeq}
\eeq
where $g$ is the determinant of metric $g_{\mu\nu}$, $|\pa \phi_2|:=\sqrt{\pa_\mu \phi_2 \pa^\mu \phi_2}$, and $P^{\mu\nu}:=g^{\mu\nu}-\pa^\mu \phi_2 \pa^\nu \phi_2/|\pa \phi_2|^2$.
Note that $\pa^\mu \phi_2 /|\pa \phi_2|$ is a spacelike unit vector, hence $P^{\mu\nu}$ represents a projection 
operator onto the $\phi_2=\text{const}$ hypersurface, which is timelike. 
Considering the property of the projection operator with $\theta=\pi/2$, it turns out that Eq.~\eqref{equatorial_waveeq} governs the wave propagation along a magnetic field line (\alf wave).
Separating the variables as $\delta \phi=\pa_\theta \phi_1 e^{-i\omega t}e^{im\vp}R(r)$ \footnote{The $\theta$-dependence of this expression originates from the definition of the perturbation. As discussed in \cite{Noda:2021ivs}, the perturbation perpendicular the magnetic surface on the equatorial plane is represented with the Lagrange displacement vector $\zeta^\mu$ as $\delta \phi:=\zeta^\mu \partial_\mu \phi_1$.}, we obtain the following differential equation for $R(r)$: 
\beq
\df{d}{dr} \kk{-\Gamma \df{dR}{dr}} + 2i {\cal{A}}  \df{dR}{dr} +\mk{i\df{d{\cal{A}}}{dr}    + U }R =0,
\label{waveeq_r}
\eeq
where $U$ and $\mathcal{A}$ are defined as
\begin{align}
   \notag U&=\sqrt{-g}|\pa \phi_2|^2 \mk{-g^{\theta \theta} -\omega^2 P^{tt}+2m \omega P^{t\vp} -m^2 P^{\vp\vp}},\\
  \notag  \mathcal{A}&=\sqrt{-g}|\pa \phi_2|^2 (mP^{r\vp}-\omega P^{rt})=J_Br^2G^t \mk{\omega- m G^\vp/G^t},
\end{align}
with $G^\vp=g^{\vp\vp}-\Omega_F g^{\vp t}$ and $G^t=g^{t\vp}-\Omega_F g^{tt}$.
As $\Gamma=0$ at $r=r_\text{in}$ and $r_\text{out}$, Eq.~\eqref{waveeq_r} has singular pgoints. However, these are just coordinate singularities and we map them to negative and positive infinity by introducing the ``tortoise'' coordinate $r_*$ as $dr_*/dr=-\Gamma^{-1}$.
Furthermore, we rescale the radial part of the wave function as $R=h(r)u(r)$ \footnote{To derive \eqref{Veff}, we eliminate 
$du/dr_*$ terms by requiring that $h(r)$ satisfy the relation $(1/h)dh/dr = i\mathcal{A}/\Gamma$, which leads to
\beq
h(r) = \exp{\kk{i \int \df{\mathcal{A}}{\Gamma}dr}}.
\eeq} and obtain the wave equation for $u$ as
\beq
\dfrac{d^2 u }{dr_*}-V_\text{eff}u=0, \quad\quad
V_\text{eff}=-\mathcal{A}^2 +\Gamma U.
\label{Veff}
\eeq
where $V_\text{eff}$ is the effective potential for \alf wave, which is shown 
in Fig.~\ref{fig:potential}.
\noindent
Since the inner light surface is a causal boundary for 
\alf waves like black hole horizon \cite{Kinoshita:2017mio, Gralla:2014yja}, 
we should impose the purely ingoing 
boundary condition on \alf waves there.
In the vicinity of the inner light surface, the effective potential asymptotically approaches a constant value $V_\text{eff}^{\text{asymp}}=-\mathcal{A}^2|_{r_\text{in}}$, and the asymptotic solutions at the inner light surface are obtained as
\beq
\label{asympt_u}
u \rightarrow  {\displaystyle\exp{\left[\pm i(\omega-m\Omega_F)  F_\text{in} \df{\Omega_\text{H}-\Omega_F}{\Omega_\text{in}-\Omega_F} \ r_* \right]}},
\eeq
where
$\Omega_\text{in}:=-g_{\vp t}/g_{\vp\vp}|_{r_\text{in}}$ is the angular velocity of ZAMO by
the frame dragging at the inner light surface, $F_\text{in}=(r_\text{in}^2/g_{\vp\vp}(r_\text{in}))(r_\text{H}^2+a^2)/r_\text{H}^2$ is positive constant, and 
$(\Omega_\text{H}-\Omega_F)/(\Omega_\text{in}-\Omega_F)$
is also positive for all $\Omega_F$ under the condition for the BZ process to be active as 
shown by the blue curve in Fig.~\ref{fig:phases}. Therefore, we should take the negative sign in the bracket in \eqref{asympt_u} for the ingoing mode.

\section{Energy extraction via \alf waves}
The essence of superradiance is that ingoing waves pass 
through the causal boundary with negative energy due to the effect of 
the frame dragging, producing the nonzero outward energy flux. The condition for superradiance is obtained by evaluating the 
relationship between the energy flux and the angular momentum flux at the causal boundary \cite{Lasota2013}. 
Since the causal boundary for \alf waves within the force-free approximation is the inner light surface, we investigate the fluxes of \alf waves there.
\subsection{Energy extraction condition in the BZ process}
 First, we review the energy extraction through the BZ process. The energy and angular momentum flux vectors are defined as $P^\mu:=-T^{\mu}_{\ \ \nu}\xi_{(t)}^{\nu}=-T^{\mu}_{\ \ t}$ and 
$J^\mu:=T^{\mu}_{\ \ \nu}\xi_{(\vp)}^{\nu}=T^{\mu}_{\ \ \vp}$, respectively, where the energy momentum 
tensor of the electromagnetic field is given by $T_{\mu\nu}= 1/(4\pi)(F_{\mu \al}F_{\nu}^{\ \ \al}-F_{\al\be}F^{ \al \be}g_{\mu\nu}/4)$.
  The radial components of the flux vectors associated with the 
  background magnetosphere \eqref{eq:EulerKerr} are obtained as
 \begin{align}
  \label{Sr}
  &P^r=-T^{r}_{\ \ t}=\df{1}{4\pi}\df{r_\text{H}^2+a^2}{r_\text{H}^2}\Omega_F (\Omega_\text{H}-\Omega_F)\df{(\pa_\theta \phi_1)^2}{r^2},\\
  \label{Lr}
&J^r=T^{r}_{\ \ \vp}=\df{1}{4\pi}\df{r_\text{H}^2+a^2}{r_\text{H}^2} (\Omega_\text{H}-\Omega_F)\df{(\pa_\theta \phi_1)^2}{r^2},
\end{align}
These fluxes are related by $P^r = \Omega_F J^r$. From \eqref{Sr}, $P^r>0$ only for 
$0<\Omega_F < \Omega_\text{H}$. 
Hence, we see that $0<P^r < \Omega_\text{H} J^r$. From the perspective of the area law of 
the Kerr spacetime \cite{Bardeen1973}, this inequality corresponds to the energy extraction 
from the black hole.
From another perspective suggested by \cite{Kinoshita:2016,Kinoshita2017}, the necessary condition for the BZ process is that the locus of the inner light surface is inside the ergoregion. This condition corresponds to both the energy flux and the angular momentum flux being positive at the inner light surface: 
$P^r|_{r_\text{in}}>0$ and $J^r|_{r_\text{in}}>0$. 
\subsection{Superradiant condition for \alf waves}
\begin{figure}[H]
 \centering
 \includegraphics[width=0.85\linewidth]{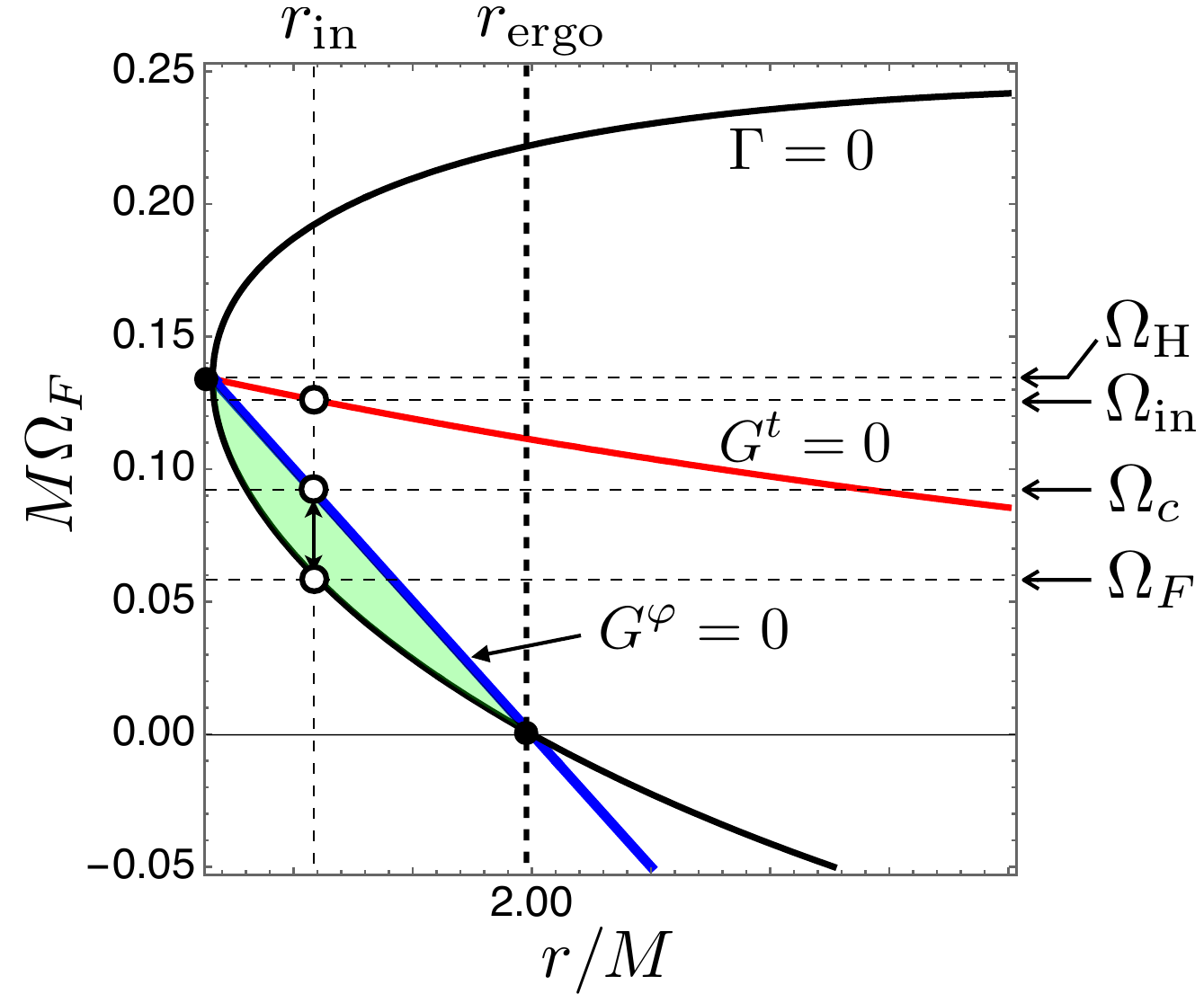}
 \caption{\footnotesize{Superradiant condition window for each given $\Omega_F$, shown with the curves of $G^\vp=0$ (blue), $G^t=0$ (red), and $\Gamma=0$ (black) in the $r\Omega_F$-plane for $a/M=0.5$. For a given $\Omega_F$, $r_\text{in}$ is determined, and hence $\Omega_F=\Omega_\text{c}$ and $\Omega_F=\Omega_\text{in}$ lines are shown as the horizontal dashed lines. The vertical double arrow in the colored region indicates the possible range of $\omega/m$ for \alfv superradiance \eqref{superrad}.}}
 \label{fig:contour}
\end{figure}

We apply the above condition for $P^r|_{r_\text{in}}$ and $J^r|_{r_\text{in}}$ to the fluxes associated with the \alf waves. Since the field strength $F_{\mu\nu}$ is written in terms of the Euler potential, the perturbed field strength due to the variation $\phi_1 \rightarrow \phi_1 + \delta \phi$ is given by 
$F_{\mu \nu} \rightarrow F_{\mu\nu} + \delta F_{\mu\nu}$.
 The contribution 
of \alf waves to the fluxes are $P_\text{Alf}^\mu=  1/(4\pi)(-\delta F^{\mu \al}\delta F_{t\al} +  \delta^{\mu}_{t} \delta F_{\al \be} \delta F^{\al \be}/4)$ and $J_\text{Alf}^\mu= 1/(4\pi)(\delta F^{\mu\al}\delta F_{\vp \al}-\delta^\mu_{\vp}\delta F_{\al \be} \delta F^{\al \be}/4)$.
To evaluate the $r$-component of these fluxes at the inner light surface, we substitute the ingoing asymptotic solution \eqref{asympt_u} and take the time average to eliminate the oscillatory terms. As a result, we obtain 
\begin{widetext}
 \begin{align}
      \label{Sralf}
  \langle P_\text{Alf}^r \rangle|_{r_\text{in}}
     &= J_B\df{ (\pa_\theta \phi_1)^2}{8\pi}m^2 \mk{\df{\omega}{m}-\Omega_F}\left.\mk{\df{\omega}{m}g^{\vp t}-g^{\vp\vp}}|u|^2\right|_{r_\text{in}},     \\
    \label{Lralf}
    \langle J_\text{Alf}^r\rangle |_{r_\text{in}}
      &=J_B\df{ (\pa_\theta \phi_1)^2}{8\pi}m^2 \mk{\df{\omega}{m}-\Omega_F}\left.\mk{\df{\omega}{m}g^{t t}-g^{t\vp}}|u|^2\right|_{r_\text{in}}. 
      \end{align}
\end{widetext}
The details of the derivation are provided in Appendix A.
Now, the condition on $\omega/m$ for the extraction of black hole's rotational energy 
is obtained from the requirements $\langle P_\text{Alf}^r\rangle|_{r_\text{in}}>0$ and $\langle J_\text{Alf}^r\rangle|_{r_\text{in}}>0$. 
Since $g^{\vp t}<0$ and $g^{tt}<0$ at the inner light surface, these 
positivity conditions yield
\begin{align}
\label{cond1}
&\mk{\df{\omega}{m}-\Omega_F}\mk{\df{\omega}{m}-\Omega_\text{c}   } < 0,\\
\label{cond2}
&\mk{\df{\omega}{m}-\Omega_F}\mk{\df{\omega}{m}-\Omega_\text{in}} < 0,
\end{align}
respectively, where $\Omega_\text{c}:=g^{\vp\vp}/g^{\vp t}|_{r_\text{in}}$
\footnote{$\Omega_\text{c}$ is defined as the value of the function $g^{\vp\vp}/g^{\vp t}$ at the inner light surface. It can be regarded as a type of angular velocity, and a similar quantity 
appears in the analysis of GRMHD accretion onto a Kerr black hole \cite{Takahashi1990}. 
When evaluated at the horizon $r_\text{H}$, this function coincides with $\Omega_\text{H}$, and it becomes zero at the ergosurface. We expect that generalizing the analysis of \alf wave propagation to the GRMHD regime shed light on physical nature of this angular velocity, and we leave this investigation for future work.}.
If there exists $\omega/m$ mode satisfying these inequalities, the rotational energy of the 
black hole is extracted by the ingoing \alf wave passing through the inner light surface. 
As shown in the Appendix B, the relationship among $\Omega_F, \Omega_\text{c},$ and $\Omega_\text{in}$ is given by $\Omega_F<\Omega_\text{c}<\Omega_\text{in}$ under the condition for the BZ process to work: $0<\Omega_F<\Omega_\text{H}$. Therefore, the inequalities \eqref{cond1} and \eqref{cond2} yield the following condition for $\omega/m$:
\beq
(0<)\ \ \Omega_F < \df{\omega}{m} < \Omega_\text{c}\ \ (<\Omega_\text{H}),
\label{superrad}
\eeq
where we put $0$ and $\Omega_\text{H}$ in the inequality to clarify the inclusion relation with the condition for the BZ process. 
The superradiant condition \eqref{superrad} indicates that when the BZ process is active, \alf waves with frequencies within the specific range \eqref{superrad} can extract rotational energy and enhance the flux through the BZ process. 
The dependence of this frequency range on $\Omega_F$ is shown in Fig.~\ref{fig:contour}. 
Since the radius of the inner light surface $r_\text{in}$ depends on $\Omega_F$, the right-hand 
side of the superradiant condition \eqref{superrad}, denoted as $\Omega_\text{c}$, is also a function of $\Omega_F$ as well. 
Therefore, once $\Omega_F$ is specified, the parameter range for 
the \alfv superradiance becomes fixed. 
To explicitly illustrate this width for a given $\Omega_F$, we also plot the curves $G^\vp=0$ and $G^t=0$, which correspond to $\Omega_F = g^{\vp\vp}/g^{\vp t}$ and 
$\Omega_F = g^{t\vp}/g^{tt}$, respectively on the $r\Omega_F$-plane. Additionally, the curve $\Gamma=0$ shows the radius of the inner light surface as the intersection with the specified $\Omega_F$. 
The vertical line at $r_\text{in}$, shown as the thin dashed line, always intersects $G^\vp=0$ and $G^t=0$. The vertical axis values at these intersection points correspond to $\Omega_\text{c}$ and $\Omega_\text{in}$, respectively. Thus, the green-shaded region is where the \alfv superradiance occurs. In Fig.~\ref{fig:contour}, the parameter range \eqref{superrad} 
for a given $\Omega_F$ is indicated by 
the double arrow. Notably, the green-shaded region is always located within the ergoregion.

\subsection{Similarity to the energy extraction via magnetohydrodynamic inflow}
Takahashi {\textit{et al.}} \cite{Takahashi1990} 
presents the condition for a negative energy magnetohydrodynamic (MHD) inflow, where the same curves $G^\vp=0$, $G^t=0$, and $\Gamma=0$ play an important role. Note that, in the MHD case, instead of the inner light surface, the inner \alf surface becomes the causal boundary for the \alf wave. Therefore, the radius of the inner light surface $r_\text{in}$ in Fig.~\ref{fig:contour} must be replaced by that of the inner \alf point $r_\text{A}$, which is one of critical points of the MHD flow. Hence, the energy extraction condition is that the inner \alf point must be located at least inside the ergosphere. The range of $\Omega_F$ to satisfy this condition is $\Omega_F^{(-)}<\Omega_F < g^{\vp\vp}/g^{\vp t}|_{r_\text{A}}$, depending on the location of the \alf point, where $\Omega_F^{(-)}$ is the minimum value of $\Omega_F$ 
for the existence of the inner light surface. This parameter region for $\Omega_F$ and $r=r_\text{A}$ corresponds 
exactly to the colored region in Fig.~\ref{fig:contour}. 
This similarity between the energy extraction through the ingoing \alf wave and the MHD inflow arises from the fact that both the \alf wave propagation (deviation of magnetic field lines) and the MHD inflow (which incorporates plasma inertia effects into the force-free magnetosphere) can be considered as perturbations to the force-free background magnetosphere. 
This prediction becomes clearer by 
calculating the energy flux using the inner \alf point instead of the light surface as the causal boundary.

\section{Effect of \alf waves on the BZ flux}
We examine the effect of \alf wave on the 
energy flux associated with the BZ process. Since we consider a magnetic surface near the equatorial plane in the force-free magnetosphere, we focus on the energy flux within a narrow latitudinal band (zonal band) around the equatorial plane. This zonal band, located at a radius $r$ around the equatorial plane, is defined as the region $\{(\theta,\vp) | \pi/2-\Delta \theta \leq \theta \leq \pi/2+\Delta \theta;\ 0 \leq \vp \leq 2\pi\}$, where $\Delta \theta \ll 1$. 
The total energy flux integrated over the zonal band at a radius  $r$ is 
denoted as $L_\text{BZ}|_{r}$ for the BZ process and $L_\text{Alf}|_{r}$ for the \alf waves.

We evaluate \alf wave flux at the inner light surface \eqref{Sralf}. Since in the original framework of the BZ process \cite{Blandford1977} the energy flux given in Eq.~\eqref{Sr} is evaluated at the horizon, we need to clarify the relationship between $L_\text{BZ}|_{r_\text{H}}$ and $L_\text{BZ}|_{r_\text{in}}$. 
By integrating \eqref{Sr} over the zonal band at an arbitrary radius $r$, we obtain
\begin{align}
\notag 
L_\text{BZ}|_{r}
&:=\int_0^{2\pi} d\vp \int_{\pi/2-\Delta \theta}^{\pi/2+\Delta \theta}d\theta \sqrt{-g} P^r \\
\label{EBZ}
&=\df{1}{4\pi} \df{r_\text{H}^2+a^2}{r_\text{H}^2} \Omega_F (\Omega_\text{H}-\Omega_F)\df{A_{\vp,\theta}^2}{r^2}\times 4\pi r^2  \Delta \theta  
\end{align}
where we expressed $\pa_\theta \phi_1=A_{\vp,\theta}$ through $F_{\mu\nu}=\pa_\mu A_\nu -\pa_\nu A_\mu$. 
Since \eqref{EBZ} is independent of $r$, it follows that 
$L_\text{BZ}|_{r_\text{H}}=L_\text{BZ}|_{r_\text{in}}$.

Similarly, integrating \eqref{Sralf} over the zonal band at the inner light surface, $L_\text{Alf}|_{r_\text{in}}$ is obtained. The total energy flux, $L_\text{tot}$, 
including the BZ process and the effect of the \alf wave is 
given by
\begin{widetext}
\beq
\label{Etot}
L_\text{tot} := L_\text{BZ}|_{r_\text{in}}+L_\text{Alf}|_{r_\text{in}}=\df{ r_\text{H}^2+a^2}{r_\text{H}^2}A_{\vp,\theta}^2\Omega_F(\Omega_\text{H}-\Omega_F)\left[ 1+ \df{ g^{\vp t}|_{r_\text{in}}}{2\Omega_F} \int d\omega \sum_m\mk{\omega- m\Omega_F} \mk{\omega - m\Omega_\text{c}} |u|^2  \right]\Delta \theta.
\eeq
\end{widetext}
The first term in the bracket is the contribution from the BZ process \cite{Blandford1977}, and the second 
term indicates the correction by the ingoing \alf wave. 
The integral over $\omega$ and the sum over $m$ are necessary to take the various modes of \alf waves into account. 
The integral and sum may be computed by analyzing the normal mode of \alf waves, but such computation lies out of the scope of the present paper. 
In the long wavelength limit ($\omega,m\rightarrow 0$), the total flux reduces to the flux of the BZ process. 
It is also important to note that while the \alf waves always propagate inward through the inner light surface toward the black hole, its contribution to the energy flux can be either positive (indicating energy extraction) or negative (indicating energy injection), depending on the wave frequency. 

\section{Discussions and Concluding Remarks}
We have investigated the energy flux of \alf waves in the force-free magnetosphere of a Kerr black hole. We derived a novel condition for superradiance via ingoing \alf waves as presented in \eqref{superrad}. This condition shows that ingoing 
\alf waves can extract energy from the Kerr black hole, depending on the wave's frequency. 
From the expression of the energy flux \eqref{Etot}, 
the energy flux of \alf wave contains 
the factor $\Omega_F(\Omega_\text{H}-\Omega_F)$ common to the 
BZ flux. Thus, we now have 
a unified understanding of the BZ flux and the \alfv superradiance. 
The unified formulation, encapsulated in \eqref{Etot}, provides an essential 
insight into the energy extraction from rotating 
black holes. It offers a conceptual framework for understanding the energy sources of relativistic jets and high-energy astrophysical phenomena that includes wave effects. This result would inform and motivate future numerical studies.

In this paper, to reveal the essence of the wave-related correction to the BZ power and for the sake of simplicity, we investigated the \alf wave propagation in the equatorial plane. We also expect that similar effects of \alf waves would arise in other magnetospheric configurations, including field lines off the equatorial plane. This is because the essential mechanism discussed in this study relies only on two key ingredients: the 
frame dragging effect of the Kerr spacetime and the presence of rotating magnetic field lines, under which \alf waves can pass through the inner light surface and fall into the black hole.

In realistic astrophysical environments around black holes, there are nonlinear effects based on plasma physics such as magnetic reconnection and the formation of plasmoids, which leads to the breakdown of the 
force-free approximation. In such cases, \alf wave propagation should be examined within 
the framework of magnetohydrodynamics.

 Furthermore, propagation of large-amplitude \alf waves is also an intriguing subject, particularly in connection with nonlinear energy transport mechanisms in black hole magnetospheres. 
 Since \alf waves in this study were introduced as a small perturbation to the background magnetosphere, the 
 contribution of the individual ($\omega,m$) modes 
 to energy extraction is expected to be minor. 
 However, analogous to the case of gravitational wave superradiance \cite{East2013}, 
 the condition for \alfv superradiance derived through linear analysis may still serve as a criterion for the energy extraction via \alf waves even in the nonlinear regime, where larger \alf wave amplitudes would be permitted.

\acknowledgements
  The authors thank Shinji Koide for fruitfull discussion.
  S.N. is supported by JSPS KAKENHI Grant No.~24K17053.

\begin{appendix}
\section{Derivation of the energy and angular momentum fluxes of \alf wave}
As discussed in Sec. III A, using the Euler potentials, 
$r$-component of 
the energy and angular momentum fluxes for \alf waves can be written as 
\begin{align}
\notag   P_\text{Alf}^r 
    &=g^{rr}\left\{-G^\vp(\delta \phi_{,r} \delta \phi_{,t} + \Omega_F \delta \phi_{,r} \delta \phi_{, \vp})\right. \\
 \label{deltaSr}   &\left. \ + \pa_r \phi_2 \kk{  g^{\vp t} \delta \phi_{,t}^2+g^{\vp\vp}\Omega_F  \delta \phi_{,\vp}^2  + (g^{\vp\vp} + \Omega_F g^{\vp t}) \delta \phi_{,t} \delta \phi_{,\vp}  } \right\},\\
  \notag J_\text{Alf}^r 
&=g^{rr} \left\{  -G^t ( \del \phi_{,r} \del \phi_{,t} +\Omega_F \del \phi_{,r} \del \phi_{,\vp} )\right.\\   
\label{deltaLr}&\ \left.+\pa_r \phi_2 
\kk{  g^{tt}\del \phi_{,t}^2 + g^{t \vp} \Omega_F \del \phi_{,\vp}^2 +(g^{t\vp} +\Omega_F g^{tt})\del \phi_{,t} \del \phi_{,\vp}   } \right\}.
\end{align}
 The fluxes need to be evaluated with real function, therefore we write $\delta \phi$ as $\delta \phi = {\pa_\theta \phi_1}/2 \mk{e^{-i\omega t} e^{im \vp} R + \text{c.c}}$
    before substituting it into the above expression. 
    Furthermore, by taking the time average, the oscillatory terms are 
    eliminated. Some terms 
    included in the expressions \eqref{deltaSr} and \eqref{deltaLr} are evaluated as
       \begin{align}
    \notag &  \langle \delta \phi_{,r} \delta \phi_{,t}   \rangle+ \Omega_F \langle \delta \phi_{,r} \delta \phi_{,\vp} \rangle \\
    \label{1} &=   (\pa_\theta \phi_1)^2\df{ (\omega -m \Omega_F)}{2} \kk{ \df{\mathcal{A}}{-\Gamma} |u|^2 + \df{1}{-2i \Gamma} W_{r_*}[u, \bar{u}]}, \\
  \notag   &g^{\vp t} \langle \delta \phi_{,t}^2 \rangle +g^{\vp\vp}\Omega_F  \langle \delta \phi_{,\vp}^2 \rangle + (g^{\vp\vp} + \Omega_F g^{\vp t})  \langle \delta \phi_{,t} \delta \phi_{,\vp} \rangle    \\
     \label{2}&= \df{ (\pa_\theta \phi_1)^2}{2}(\omega- m\Omega_F) (\omega g^{\vp t} - m g^{\vp\vp})|u|^2,\\
   \notag & g^{tt} \langle \delta \phi_{,t}^2 \rangle +g^{t\vp}\Omega_F  \langle \delta \phi_{,\vp}^2 \rangle + (g^{t\vp} + \Omega_F g^{t t})  \langle \delta \phi_{,t} \delta \phi_{,\vp} \rangle\\
    \label{3}&= \df{ (\pa_\theta \phi_1)^2}{2}(\omega- m\Omega_F) (\omega g^{t t} - m g^{t\vp})|u|^2,
   \end{align} 
   where $W_{r_*}[u,\bar{u}]=u d\bar{u}/dr_* - \bar{u} du/dr_*$, and the symbol
   $\langle \ \cdot\  \rangle$ represents the time-averaged quantity. 
   Note that \eqref{1} seems to diverge at the inner light surface ($\Gamma=0$), however by substituting the ingoing mode, which is given by \eqref{asympt_u} with negative sign, the divergence can be removed.
   Finally, the fluxes at the inner light surface are obtained as \eqref{Sralf} and \eqref{Lralf}.
\section{Derivation of the relation among $\Omega_F, \Omega_\text{c}$, and $\Omega_\text{in}$}
The inequality $\Omega_F< \Omega_\text{c}<\Omega_\text{in}$ plays a crucial role in deriving the superradiant condition \eqref{superrad}. In this appendix, we provide a derivation of this inequality. 
Since this study focuses on the correction to the BZ power, we consider the regime in which 
the BZ process is active,i.e., $0<\Omega_F < \Omega_\text{H}$. 
Under this condition, we first note that $\Omega_F<\Omega_\text{in}$ as illustrated in Fig.~\ref{fig:phases}. Using this fact and the relation below, we establish that $\Omega_F < \Omega_\text{c}$:
\begin{align}
\notag \Omega_\text{c}-\Omega_F&=\left.\df{g^{\vp\vp}-\Omega_F g^{t\vp}}{g^{t\vp}}\right|_{r_\text{in}}\\
\notag
&=\left.\df{\Omega_F(g^{t\vp}-\Omega_F g^{tt})}{g^{t\vp}}\right|_{r_\text{in}}\\
\notag
&=\left.\Omega_F \df{g^{tt}}{g^{t\vp}}\mk{\df{g^{t\vp}}{g^{tt}}-\Omega_F }\right|_{r_\text{in}}\\
&=\df{\Omega_F}{\Omega_\text{in}}(\Omega_\text{in}-\Omega_F)>0,
\end{align}
where in the second equality, we have used the condition 
$\Gamma(r_\text{in})=0$. 
Furthermore, since $\Omega_F/\Omega_\text{in}<1$, we obtain the inequality $\Omega_\text{c}-\Omega_F < \Omega_\text{in}-\Omega_F$, which leads to $\Omega_\text{c} < \Omega_\text{in}$. 
To summarize, we arrive at the following ordering:
\beq
\Omega_F < \Omega_\text{c} < \Omega_\text{in}.
\eeq

\end{appendix}

\bibliography{main}
\bibliographystyle{JHEPmod}
\end{document}